\documentclass[twocolumn,twoside,slac_two]{revtex4}
\usepackage{graphicx}
\usepackage{fancyhdr}
\begin{document}

\title{Hunting for GeV emission from the $\gamma$-ray binary HESS J0632+057}

%

\author{A. B. Hill}
\affiliation{W. W. Hansen Experimental Physics Laboratory, KIPAC, Department of Physics and SLAC National Accelerator Laboratory, Stanford University, Stanford, CA 94305, USA }
\affiliation{Faculty of Physical \& Applied Sciences, University of Southampton, Southampton, SO17 1BJ, UK }
\author{A. Caliandro}
\affiliation{Institut de Ciencies de l'Espai (IEEC-CSIC), Campus UAB, 08193 Barcelona, Spain}
\author{on behalf of the Fermi Large Area Telescope Collaboration}

\begin{abstract}
In the last decade Cherenkov telescopes on the ground and space-based $\gamma$-ray instruments have identified a new sub-class of high mass X-ray binaries (HMXB), whose emission is dominated by $\gamma$ rays.ÊTo date only five of these systems have been definitively identified.  However at GeV energies there is still one, HESS J0632+057, that has no reported detection by the Fermi LAT.Ê A deep search for $\gamma$-ray emission of HESS J0632+057 has been performed using more than 3.5 years of Fermi-LAT data. We present the results of this search in the context of the other known $\gamma$-ray binary systems.

\end{abstract}

\maketitle

\thispagestyle{fancy}


\section{Introduction}
The population of $\gamma$-ray binaries comprises of a handful of high mass X-ray binaries that have been detected at high (0.1--100 GeV) or very high ($>$100 GeV) energies with the peak of their emission lies within the  $\gamma$-ray band \citep[see e.g.][]{ref1}.  Five such systems are currently known: LS~I~+61$\circ$303; LS 5039; PSR B1259$-$63; 1FGL J1018.6$-$5856; HESS J0632+057. In all but one of  these systems the nature of the compact object is unknown and consequently the origin of the high energy emission is still not clear; the exception is PSR B1259$-$63 which is known to host a radio pulsar.

\begin{figure}[h,b,t]
\centering
\includegraphics[width=82mm]{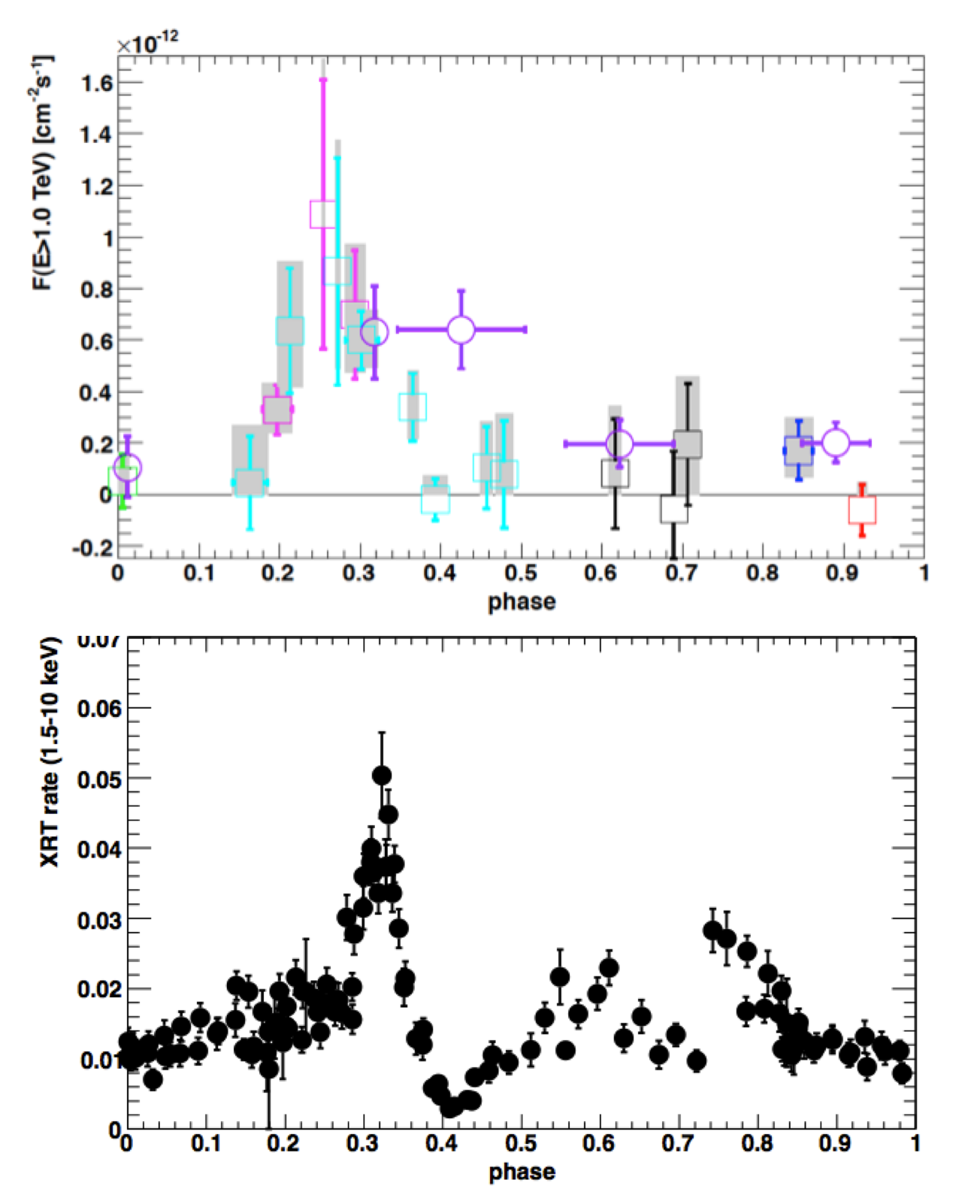}
\caption{\emph{Top}: The TeV phase folded light curve for HESS J0632+057 from H.E.S.S. (circles; 2004 \& 2010) and VERITAS (squares; 2006 to 2011) \citep{ref4}. \\ \emph{Bottom}: The Swift-XRT 1.5--10 keV phase folded light curve \citep{ref4}.}\label{FigLC}
\end{figure}

In 2007 the H.E.S.S. collaboration reported the discovery of a new TeV point-source lying in the Galactic Plane, HESS J0632+057 \citep{ref2}.  Within the H.E.S.S. error circle was a massive emission-line star, MWC 148, of spectral type B0pe which pointed to the possible binary nature of the object.  Subsequent multi-wavelength follow up identified a periodicity in the X-ray and  $\gamma$-ray at 321 $\pm$ 5 days \citep{ref3,ref4} confirming the binary nature of this object and adding a new member to the class of  $\gamma$-ray binaries; see Figure~\ref{FigLC}.  The MAGIC team recently reported the first detection of  $\gamma$-ray emission in the 136--400 GeV range \citep{ref5}.

\begin{figure*}[t]
\centering
\includegraphics[width=160mm]{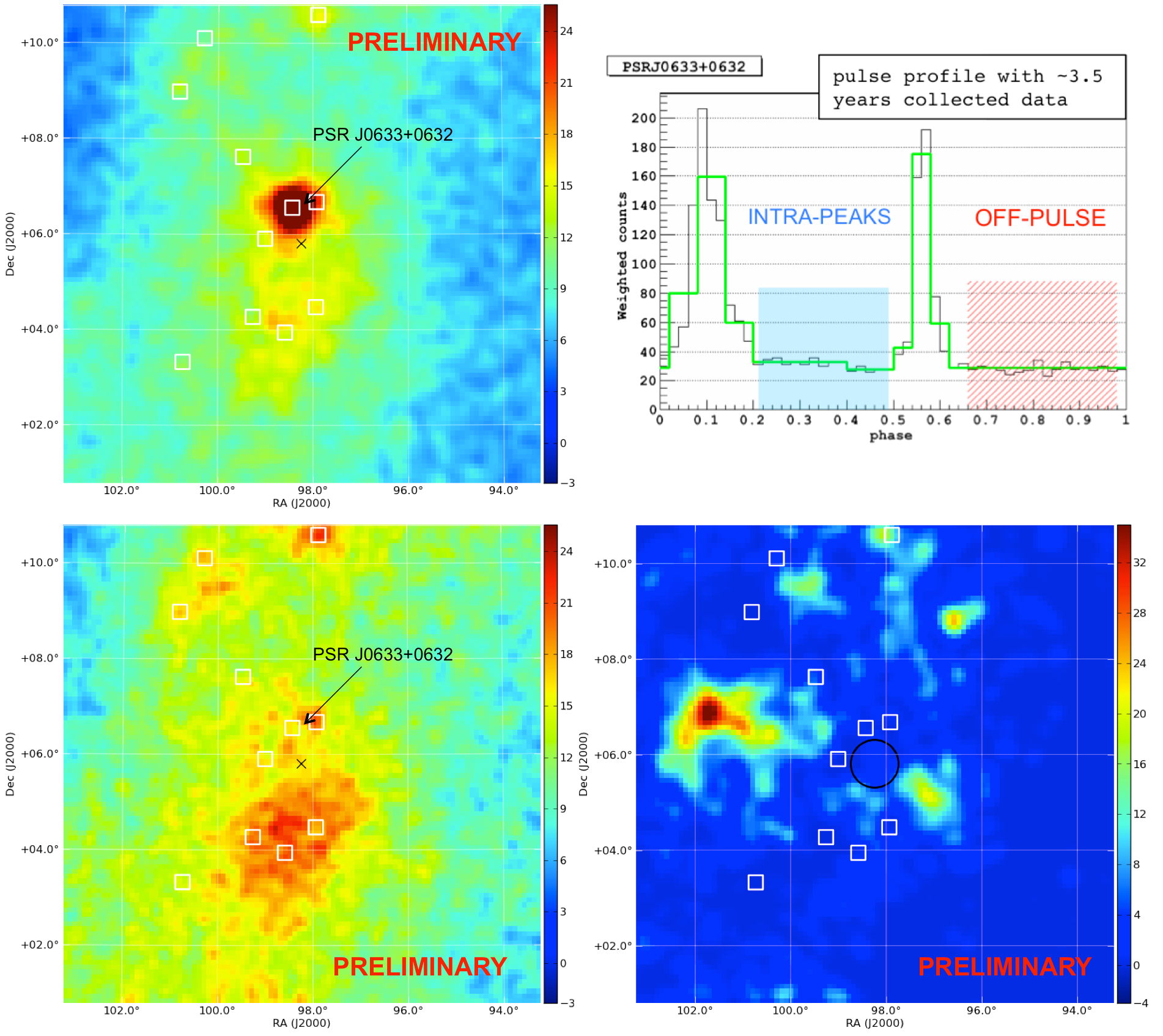}
\caption{The white squares indicate the location of 2FGL catalog sources and the black cross indicates the known position of HESS J0632+057.
\emph{Top left}: Smoothed LAT counts map above 100 MeV of the region centred on the location of HESS J0632+057. 
\emph{Top right}: The phase folded light curve of PSR J0633+0632.
\emph{Bottom left}: The smoothed LAT counts map of the region after excluding the pulsed emission of PSR J0633+0632. This map is rescaled to have equivalent exposure to the top counts map.
\emph{Bottom right}: The residual TS map having fit the LAT data with a model comprising only of 2FGL catalog sources together with a Galactic diffuse and isotropic diffuse component.
} \label{FigCtsMap}
\end{figure*}

\section{The GeV environment around \\HESS J0632+057}
In searching for emission from HESS J0632+057 at GeV energies we examined ~3.5 years of Fermi LAT data at energies $>$100 MeV using `source' class event data and the \textsc{P7SOURCE\_V6} instrument response functions.  The analysis is challenging due to the location of the source in an area of high Galactic diffuse emission and the close proximity of a bright  
$\gamma$-ray pulsar, PSR J0633+0632, $\sim$0.8$^\circ$ away from the nominal location of HESS J0632+057.  This is indicated in the smoothed LAT counts map shown in the top left of Figure~\ref{FigCtsMap}.

To minimise contamination from the pulsar we apply the Ôpulsar gatingÕ procedure of only selecting events which occur outside of the two peaks of PSR J0633+0632.  To define the `unpulsed' region of the pulse profile we subdivide the profile into `Bayesian Blocks' \citep{1998ApJ...504..405S} and then choose the lowest blocks between the peaks.  Using only events from the intra-peak and off-pulse regions can be seen to drastically reduce the pulsar emission in the region; see the upper right and lower left panels of Figure~\ref{FigCtsMap} for the selected phases of the pulse profile and the new `pulsar gated' counts map of the region.

Fitting the Ôpulsar gatedÕ events with a model based upon the 2FGL catalog \citep{ref6} and the Galactic and isotropic diffuse models indicates that there are a number excesses in the TS map which are not accounted for; see the TS map in the lower right panel of Figure~\ref{FigCtsMap}. 

\begin{table*}\caption{A comparison of the general high-energy properties of the confirmed HMXBs detected in $\gamma$-rays .  Persistent HE sources are quoted with mean fluxes in the LAT energy band while transient sources are specified by a flux range.}
\begin{center}
\begin{tabular}{lcccccccc}
\hline
Source & Companion & Distance & Orbital period & Persistent/ & F$_{0.1-100\,{\rm GeV}}$  \\
 & sp. type & kpc & days & Transient & $10^{-10}\,{\rm erg}\,{\rm cm}^{-2}\, {\rm s}^{-1}$ \\
\hline
LS~I~+61$^\circ$303$^{a}$ & B0Ve & 2 & 26.496 & Persistent & 5.0$^h$   \\
LS~5039$^{b}$ & O6.5V((f)) & 2.5 & 3.90603 & Persistent & 2.9$^h$\\
PSR B1259$-$63$^{c}$ & O9.5Ve & 2.3 & 1236.79 & Transient & (0.9--4.4)$^{\ddag}$$^{i}$\\
1FGL J1018.6$-$5856$^{d,e}$ & O6V((f)) & 5.4 & 16.58 & Persistent & $2.8$$^d$\\
HESS J0632+057$^{f}$ & B0Vpe & 1.4 & 321 & -- & $<0.3$\\
\hline
\end{tabular}
\item $^{\ddag}$ Range given due to the transient/flaring nature of the source.\\
$^a$\,\cite{1998MNRAS.297L...5S};  $^b$\,\cite{2005MNRAS.364..899C};  $^c$\,\cite{2011ApJ...732L..11N};  $^d$\,\cite{1FGLbinary_science}; $^e$\,\cite{2011PASP..123.1262N}; $^f$\,\cite{casares2012}; $^h$\,\cite{2012ApJ...749...54H}; $^i$\,\cite{B1259}
\end{center}
\label{binary_pop}
\end{table*}

\section{Modelling the local environment of the source}
It is clear from the remaining residuals in the TS map that additional sources need to be incorporated into the model of the region.  Initially if residuals are found at or near the position of known 2FGL sources then they are re-localized and their flux normalizations allowed to vary when fitting the data.  Additional residuals must be accommodated by adding new sources to the model; this is performed in an iterative fashion to minimise the number of new sources added to the model. The process consisted of the following stages:
\begin{itemize}
 \item[-] Calculate the residual TS map using the current spectral-spatial model.
 \item[-] Add an additional point source to the model at the location of the highest excess in the TS map, modelled by a simple power-law.
 \item[-] Fit the best position of the additional source with the \textsc{pointlike} tool.
 \item[-] Re-fit with {\tt gtlike} the spectra of all the sources in the region, setting free also the spectral parameters of the additional source.
\end{itemize}

\begin{figure}[h,b,t]
\centering
\includegraphics[width=82mm]{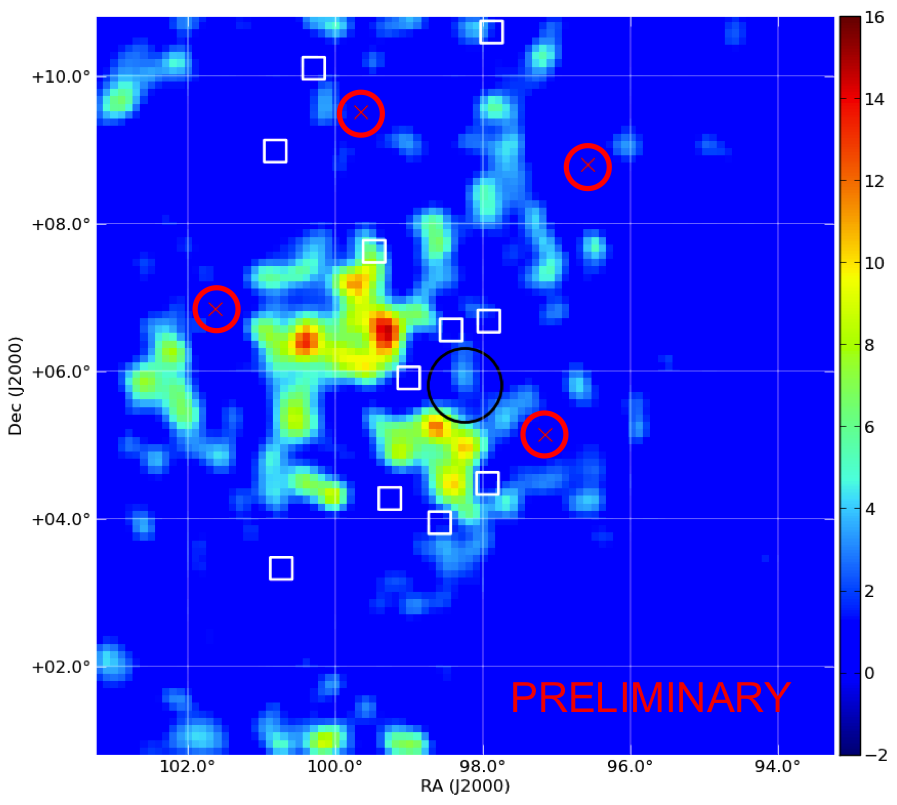}
\caption{The final residual TS map having iteratively added additional sources into the source model.  The white squares indicate the location of 2FGL catalog sources, the black circle indicates the known position of HESS J0632 +057 and the red circles with crosses indicate new non-2FGL sources.  The addition of 4 additional sources is sufficient to reduce the residuals to have TS$<$16.
.}\label{FigFinalField}
\end{figure}

The addition of four new sources is sufficient to reduce the residuals in the $>$100 MeV TS map to be below the $\sim$3$\sigma$ level.  The final residual TS map is shown in Figure~\ref{FigFinalField} and indicates the locations of the four additional sources.  None of the new sources is located at a position consistent with HESS J0632+057 and there is no indication of significant residual emission at the known location of the binary. 

\section{HESS J0632+057: \\The missing $\gamma$-ray binary}
Our analysis shows no indication of a Fermi LAT detection of persistent emission from the $\gamma$-ray binary HESS J0632+057.  Furthermore a search for emission at different orbital phases or for flaring behaviour from the source yields no significant detections.  Some of the properties of the known $\gamma$-ray binaries are shown in Table~\ref{binary_pop}.  It is clear that the $\gamma$-ray binary population spans a wide range of orbital periods and companion types, however, HESS J0632+057 clearly stands out in that despite being the nearest of these systems to Earth, if there is isotropic emission at GeV energies it is much weaker than any other identified systems and consequently it remains the only member of the population to not yet be detected at GeV energies by the LAT.

\section{Conclusions}
There is no significant detection of HESS J0632+057 at GeV energies when using 3.5 years of Fermi-LAT data.  A search for emission at different orbital phases also does not yield a detection and regular monitoring has not identified any flaring behaviour down to week timescales.  As a consequence HESS J0632+057 remains the only known $\gamma$-ray binary to not be detected by the LAT. For the energy range 0.1--100 GeV we infer a 95\% confidence flux upper limit for the source of $<$3 $\times 10^{-8}$ ph cm$^{-2}$ s$^{-1}$.

\bigskip 
\begin{acknowledgments}
A.~B. Hill acknowledges that this research was supported by a Marie Curie International Outgoing Fellowship within the 7th European Community Framework Programme (FP7/2007--2013) 
under grant agreement no. 275861. 

The \textit{Fermi} LAT Collaboration acknowledges generous ongoing support
from a number of agencies and institutes that have supported both the
development and the operation of the LAT as well as scientific data analysis.
These include the National Aeronautics and Space Administration and the
Department of Energy in the United States, the Commissariat \`a l'Energie Atomique
and the Centre National de la Recherche Scientifique / Institut National de Physique
Nucl\'eaire et de Physique des Particules in France, the Agenzia Spaziale Italiana
and the Istituto Nazionale di Fisica Nucleare in Italy, the Ministry of Education,
Culture, Sports, Science and Technology (MEXT), High Energy Accelerator Research
Organization (KEK) and Japan Aerospace Exploration Agency (JAXA) in Japan, and
the K.~A.~Wallenberg Foundation, the Swedish Research Council and the
Swedish National Space Board in Sweden.  Additional support for science analysis during the operations phase is gratefully
acknowledged from the Istituto Nazionale di Astrofisica in Italy and the Centre National d'\'Etudes Spatiales in France.

Work supported by Department of Energy contract DE-AC03-76SF00515.
\end{acknowledgments}

\bigskip 

\end{document}